\documentclass[aps,pra,twocolumn,groupedaddress,floatfix,showpacs]{revtex4}
\usepackage{amssymb,amsmath}
\usepackage{graphicx}
\usepackage{subfigure}
\usepackage[english]{babel}
\usepackage{float}
\usepackage{color}

\begin{document}

\title{Perfect localization on flat band binary one-dimensional photonic lattices}

\author{Gabriel C\'aceres-Aravena and Rodrigo A. Vicencio}

\address{Departamento de F\'{\i}sica and Millennium Institute for Research in Optics (MIRO), Facultad de Ciencias F\'isicas y Matem\'aticas, Universidad de Chile, Santiago, Chile}


\begin{abstract}
The existence of flat bands is generally thought to be physically possible only for dimensions larger than one. However, by exciting a system with different orthogonal states this condition can be reformulated. In this work, we demonstrate that a one-dimensional binary lattice supports always a trivial flat band, which is formed by isolated single-site vertical dipolar states. These flat band modes correspond to the highest localized modes for any discrete system, without the need of any aditional mechanism like, e.g., disorder or nonlinearity. By fulfilling a specific relation between lattice parameters, an extra flat band can be excited as well, with modes composed by fundamental and dipolar states that occupy only three lattice sites. Additionally, by inspecting the lattice edges, we are able to construct analytical Shockley surface modes, which can be compact or present staggered or unstaggered tails. We believe that our proposed model could be a good candidate for observing transport and localization phenomena on a simple one-dimensional linear photonic lattice.
\end{abstract}

\maketitle

\section{Introduction}

The propagation of waves in periodical systems are the natural framework to explore transport and localization phenomena in diverse fields of physics~\cite{rep1,rep2,PT}. For example, the first experimental observation of Anderson-like localization in disordered linear systems~\cite{anderson} was made in 2007, in two-dimensional (2D) induced photonic lattices~\cite{moti} and, subsequently, in fabricated one-dimensional (1D) waveguide arrays~\cite{mora}. More recently, an important theoretical and experimental interest on flat band (FB) systems has emerged~\cite{BWB08,FBluis,flach1,flach2}, showing interesting localization and transport properties on linear lattices. The current experimental techniques allow direct and indirect excitation of flat band phenomena~\cite{liebprl,liebseba,chen1,chen2,OLsaw,OLdiamond,StubSR,graphenNJP,solidstate,BECStan,pillarsAmo,acustic}, which is associated to destructive interference on specific lattice geometries. Specificaly, a Lieb photonic lattice was chosen to demonstrate, for the first time in any physical system, the existence of FB localized states~\cite{liebprl,liebseba}. A FB lattice geometry allows a precise cancellation of amplitudes outside the FB mode area, what effectively cancels the transport of energy across the system. Flat-band systems possess a linear spectrum where at least one band is completely flat or thin compared to the next energy gap. This implies the need of having a system with an unit cell composed of at least two sites and, therefore, at least two bands~\cite{FBluis,prbpal}. In general, light propagating in FB lattices will experience zero or very low diffraction, when exciting some specific sites at the unit cell~\cite{NJPlieb,OLsaw}. 

A very interesting feature of flat band systems is the possibility to construct highly localized eigenmodes by means of a destructive linear combination of extended linear wave functions~\cite{BWB08,linePRL}. These FB states are spatially compact, occupy only few lattice sites, and rapidly decay to a completely zero tail as soon as a destructive interference condition is fulfilled~\cite{FBluis}. This is a very remarkable property because FB lattices naturally generate localized structures in a linear regime, with a localization length of the order of a single unit cell. Moreover, as these linear localized modes posses all the same frequency, they are completely degenerated and any linear combination of them will also be a stable propagating solution. This can be used to achieve a non-diffractive transmission of optically codified information~\cite{liebprl,kagsignals,StubSR,linePRL,graphenNJP,zhensignal1,zhensignal2}.

Almost all studies on photonic lattices have considered single-mode excitation only. This has been reinforced due to the experimental complexity in the excitation of higher-order modes on a given lattice system, which has found a partial solution only by the implementation of a selective p-band population in cold-atom systems~\cite{becdip1,becdip2} and micropillars arrays~\cite{amo1}. But, optical waveguides can also host higher order modes, depending on the specific experimental parameters used to perform the experiment (waveguide arrays are typically fabricated considering single-mode waveguides at a given wavelength; however, by reducing this parameter higher-order modes can be excited as well). Their excitation could promote richer dynamics and new interesting phenomena, as it has been suggested for cold-atoms loaded in optical potentials~\cite{becdip3,becdip4,becdip5,becdip6}. However, a precise excitation of dipolar states has been possible very recently in optical waveguide lattices by using an image generator setup based on spatial light modulators~\cite{graphenNJP,dipole1D}. There, a well-defined contrast between the transport of fundamental and dipolar states has been shown clearly. The possibility to experimentally excite and control higher bands excitations, in optical lattice systems, paves the venue in which the study of remarkable properties of correlated systems such as superfluidity, superconductivity, organic ferromagnetic, antiferromagnetic ordering, among others, could become a concrete possibility~\cite{becdip3,becdip4,becdip5,becdip6,dipoplus1}.

As it is well known, diverse interactions have been proposed during several years to achieve stable energy localization on a lattice. For example, disorder~\cite{moti,mora}, impurities and defects~\cite{chen3,chen4}, or even nonlinearity~\cite{PRL98,PRL03,OE05}. However, all these mechanisms necessarily destroy the periodicity of the system, what finally has important consequences on the transport of energy across the lattice. In the present work, we propose a new model for the observation of FB properties. We focus on a binary 1D lattice, which to our knowledge corresponds to the simplest physical configuration for studying FB localization. We specifically concentrate on proposing a simple -- completely periodic -- system which could show the conditions to observe localization and transport of energy on a linear regime. Our model possesses a trivial FB which corresponds to an effectively isolated dipolar mode. This mode is localized at a single waveguide corresponding, therefore, to the most localized FB state ever. By precisely tuning the model parameters, we observe that a second FB can be excited as well, with states occupying only three lattice sites. In addition, we explore edge localization and find analytical Shockley edge modes with different decaying properties. At the end, we find an additional flat band when assuming equal propagation constants on both orthogonal states.

\begin{figure}[t!]\centering
\includegraphics[width=8.8cm]{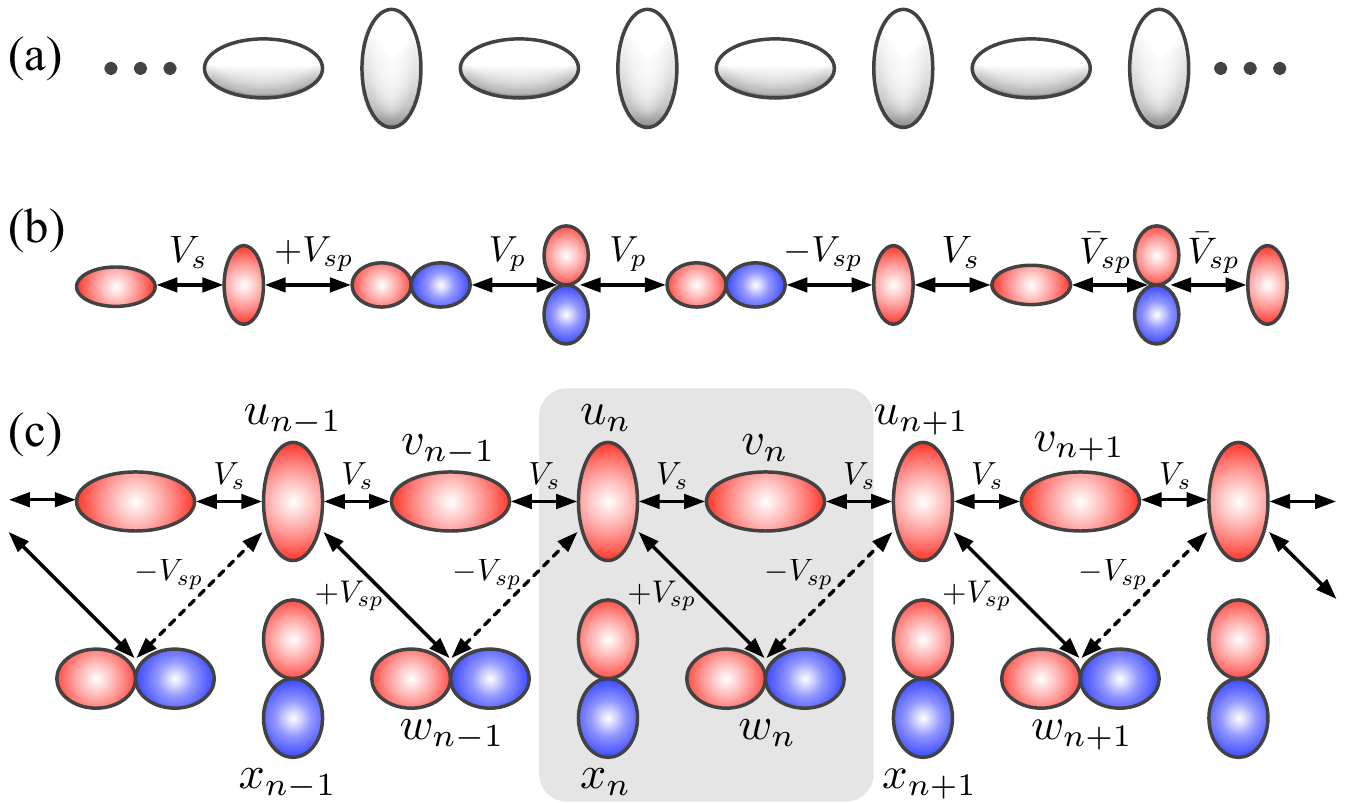}
\caption{(a) 1D-binary waveguide array. (b) Coupling interactions for this model (red represents a positive amplitude while blue a negative one). (c) Effective ribbon lattice when considering two modes per waveguide (the shaded area indicates the unit cell).}\label{fig1}
\end{figure}

\section{Model}

We study the propagation of light, in weakly coupled waveguide arrays, based on a coupled mode approach which originates from a paraxial wave equation and ends up with a set of Discrete Lineal Schr\"odinger-like equations~\cite{rep1,rep2}. This approach assumes an evanescent interaction between the modes of neighboring waveguides, with a coupling coefficient defined via the superposition integral between both mode wave-functions. Obviously, this interaction is negligible when waveguides are far away in distance and becomes physically relevant only when waveguides are close enough. Typical experiments on laser-written photonic lattices~\cite{guiasalex} define a distance of around $20$ microns to correctly describe the system assuming nearest-neighbors interactions only. In this work, we model a binary one-dimensional photonic lattice composed of an alternated configuration of waveguides, as shown in Fig.~\ref{fig1}(a). We assume elliptically oriented waveguides, which are the standard geometry in laser-written systems~\cite{guiasalex}, with the propagation coordinate $z$ (dynamical variable) running perpendicular to the transversal waveguide profile. Additionally, we consider that each waveguide supports only two orthogonal modes, the fundamental ($s$) and the dipolar ($p$) ones. In general, a single waveguide possesses always at least one bound state, which corresponds to the symmetric $s$ mode~\cite{snyder,gloge}. However, depending on the experimental conditions, it is possible to directly excite higher-order states as, e.g., $p$ modes~\cite{dipole1D,graphenNJP}. For a given waveguide, having a defined geometry and refractive index contrast, the excitation wavelength can be tuned experimentally to excite higher-order states. As different modes have a different spatial configuration, there will be a natural mismatch in their propagation constants. This implies that $\beta_s\neq \beta_p$~\cite{graphenNJP}, where $\beta_i$ is the longitudinal propagation constant of the $i$-mode at any lattice waveguide. 

The possible interactions between modes at different neighboring waveguides are depicted in Fig.~\ref{fig1}(b). Considering the symmetry of $s$ and $p$ wave-functions, we construct a general interaction rule for our binary system: the coupling between $s$ modes (defined as $V_s$) is always positive; the coupling between $p$ modes ($V_p$) as well as the coupling between $s$ and vertical $p$ modes ($\bar{V}_{sp}$) are always zero; the coupling between vertical $s$ and horizontal $p$ modes ($V_{sp}$) is defined positive when the $s$ mode is at the left-hand side, if not a minus sign is applied. In general, due to the larger area occupied by $p$ modes, $V_{sp}>V_s$. Having this in mind, we construct an effective ribbon lattice in Fig.~\ref{fig1}(c) and write the effective dynamical equations as follows
%
\begin{eqnarray}
\begin{split}
 -i\frac{\partial u_n(z)}{\partial z} &= \beta_s u_n+V_s (v_{n}+v_{n-1}) + V_{sp} (w_n-w_{n-1})\ ,\\ 
-i\frac{\partial v_n(z)}{\partial z} &= \beta_s v_n+V_s (u_{n+1}+u_{n}) \ ,\\
-i\frac{\partial x_n(z)}{\partial z} &= \beta_p x_n\ ,\\ 
-i\frac{\partial w_n(z)}{\partial z} &= \beta_p w_n-V_{sp} (u_{n+1}-u_n)\ .\label{eqs}
\end{split}
\end{eqnarray}
Here, $u_n$ and $v_n$ ($x_n$ and $w_n$) describe the amplitude of fundamental (dipolar) modes at the $n$th unit cell. The alternated orientation of our 1D binary lattice and the possibility of exciting two modes per waveguide generate a four-state effective system, which is described by these four coupled equations. It is important mentioning that, in order to have an effective dynamical interaction between the $s$ and the $p$ modes, $\Delta \beta\equiv \beta_s-\beta_p$ has to be of the order of $V_{sp}$. If not, this detuning effectively decouples the interaction between these two modes and they simply do not interact~\cite{graphenNJP}.

\section{Linear spectrum}

\begin{figure}[t!]\centering
\includegraphics[width=8.6cm]{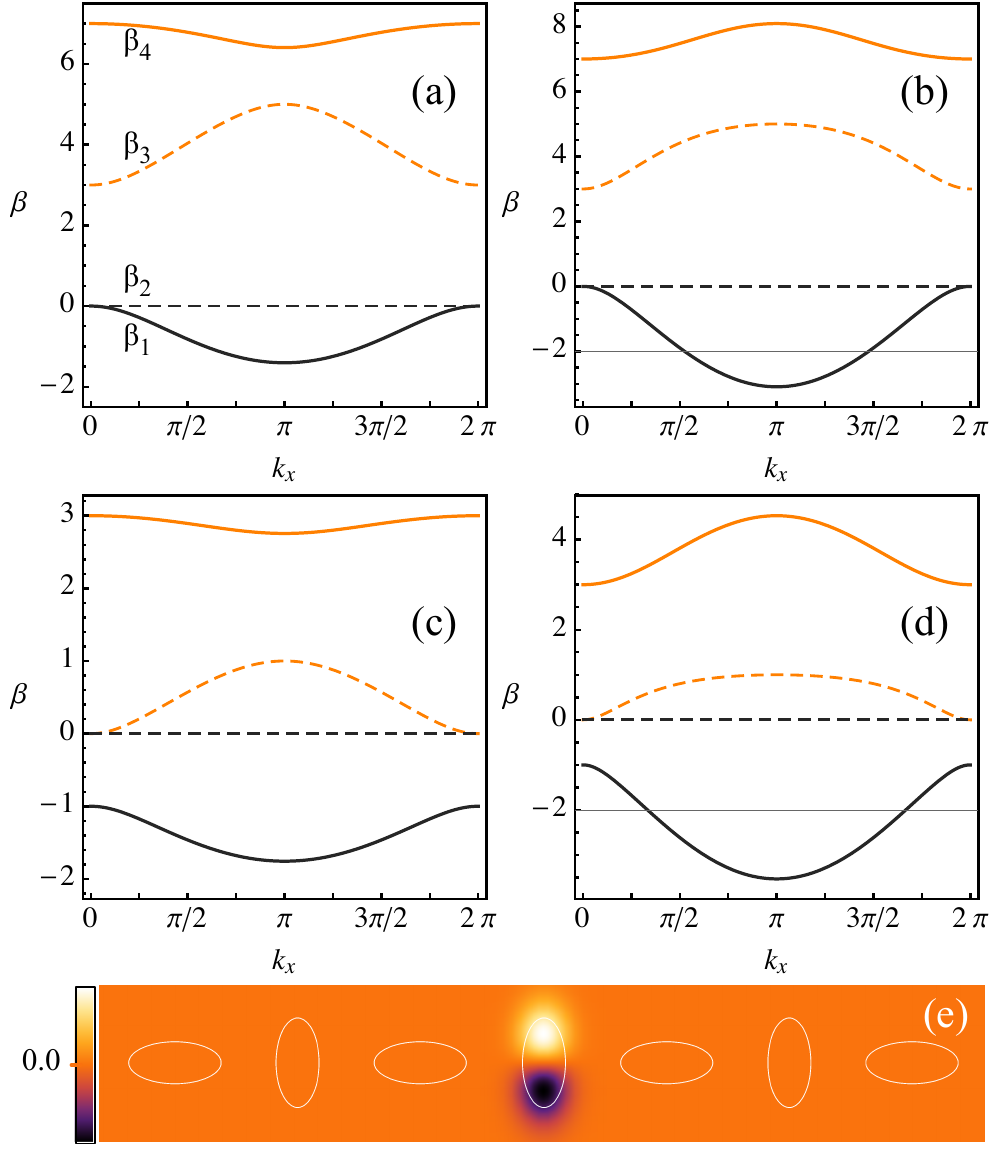}
\caption{Linear spectrum for the 1D binary lattice. (a) $V_{sp}=1.5$ and (b) $2.5$, for $\Delta\beta=5$. (c) $V_{sp}=1.1$ and (d) $2.0$, for $\Delta\beta=1$. Full black, dashed black, dashed orange, and full orange correspond to $\beta_1$, $\beta_2$, $\beta_3$ and $\beta_4$ bands. (e) FB mode amplitude profile at $\beta_2=0$. In all the figures, we set $V_s=1$.}\label{fig2}
\end{figure}
%
We solve the linear spectrum of the system by inserting into (\ref{eqs}) a standard plane-wave ansatz~\cite{FBluis}, of the form
\begin{equation*}
\{u_n,v_n,x_n,w_n\}(z)=\{A,B,C,D\}\ e^{i k n} e^{i (\beta+\beta_p) z}\ .
\end{equation*}
With this, we assume that the wave propagation occurs along the horizontal direction only, being $k$ equal to the normalized transversal wavector. $\beta$ represents the longitudinal propagation constant of the lattice eigenmodes (also known as supermodes), while $\beta_i$ represents the longitudinal propagation constant of mode $i$ on a single waveguide. Without loss of generality, we include a gauge transformation on $\beta_p$, with the purpose of reducing the model parameters and simplify the overall description. By doing this, we get a set of four coupled equations which can be written as follows
\begin{eqnarray}\label{ep}
\beta \Psi =\hspace{0.cm}\\ \begin{pmatrix}
   \Delta \beta   & V_s(1+e^{-i k}) & 0 & V_{sp}  (1-e^{-i k})  \\
     V_s(1+e^{i k}) & \Delta \beta   & 0 & 0 \\
        0   & 0 & 0 & 0\\
           V_{sp}  (1-e^{i k})  & 0 & 0 & 0
\end{pmatrix} \Psi\ , \nonumber
\end{eqnarray}
with $\Psi\equiv\{A,B,C,D\}$. By solving this eigenvalue problem, we obtain four solutions, with three of them being determined by the following third-order equation $$\left[\beta(\Delta \beta-\beta)^2-4V_s^2 \beta \cos^2\bar{k}+4V_{sp}^2(\Delta \beta-\beta) \sin^2 \bar{k}\right]=0\ ,$$ where $\bar{k}\equiv k/2$. These three bands are analytically non trivial and have no a simple and compact form for arbitrary parameters. Therefore, we show them graphically only in Figs.~\ref{fig2}(a)--(d), in the first Brillouin zone, using full black and orange lines. We also show in these figures (using dashed black lines) a completely constant solution $$\beta_2 =0\ ,$$ which corresponds to the lattice second band, as defined below. This trivial and completely flat band is related to the excitation of isolated vertical $p$ modes only. As these modes possess no coupling at all with nearest-neighbor waveguides, once they are excited they remain localized at the input position as long as the system length. This can be understood easily by directly integrating the third equation in (\ref{eqs}), getting $x_n(z)=x_n(0)\exp \{i\beta_p z\}$; being, therefore, a trivial stationary solution. In Fig.~\ref{fig2}(e) we show a sketch of this FB state. This corresponds to the most localized FB state ever, which occupies only one site of the lattice. As this mode can be excited in every vertically oriented waveguide (using arbitrary amplitudes), this trivial band can be used, for example, to transmit optically codified information through this 1D lattice.

\begin{figure}[t!]\centering
\includegraphics[width=8.6cm]{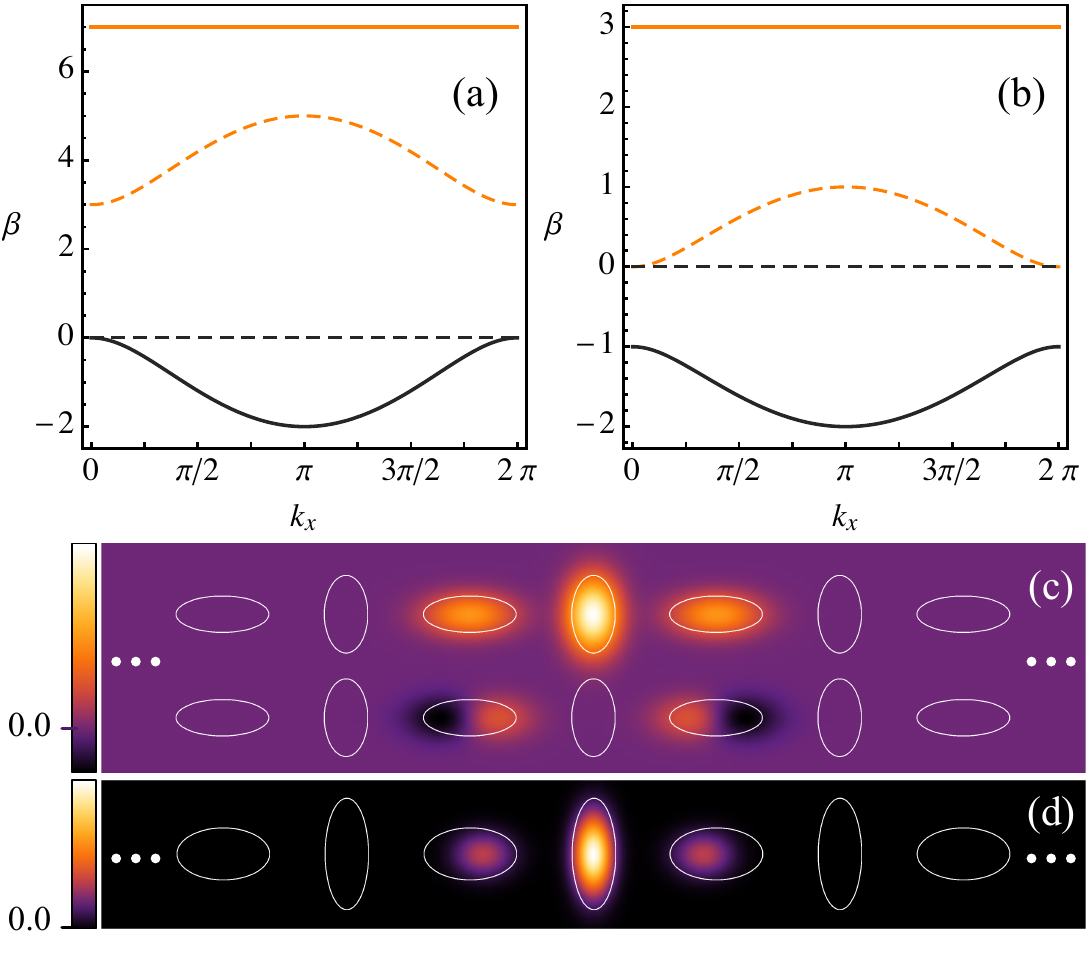}
\caption{Linear spectrum for (a) $\{\Delta\beta,V_{sp}^{FB}\}=\{5,1.87\}$ and (b) $\{\Delta\beta,V_{sp}^{FB}\}=\{1,1.22\}$. Full black, dashed black, dashed orange, and full orange correspond to $\beta_1$, $\beta_2$, $\beta_3$ and $\beta_4$ bands, respectively. (c) Effective amplitude and (d) intensity FB ($\beta_4$) mode profiles for $\Delta\beta=1$. In all the figures, we set $V_s=1$.}\label{fig3}
\end{figure}

We observe that the linear spectrum is quite symmetric; therefore, we analyze the four linear bands considering an increasing order denoted by $\beta_1$, $\beta_2$, $\beta_3$ and $\beta_4$, as indicated in Fig.~\ref{fig2}(a). At $k_x=0$, band edges become $$\{0,0,\Delta\beta-2V_s,\Delta\beta+2V_s\}\ ,$$
for $\Delta\beta\geqslant2V_s$, and $$\{\Delta\beta-2V_s,0,0,\Delta\beta+2V_s\}\ ,$$ for $\Delta\beta<2V_s$. At $k_x=\pi$, band edges are always $$\left\{\frac{\Delta\beta-\sqrt{\Delta\beta^2+16V_{sp}^2}}{2},0,\Delta\beta,\frac{\Delta\beta+\sqrt{\Delta\beta^2+16V_{sp}^2}}{2}\right\},$$
as shown for some specific parameters in Figs.~\ref{fig2}(a)--(d).
First of all, we notice that there is no gap between bands $\beta_1$, $\beta_2$ and $\beta_3$ for $\Delta\beta=2V_s$. Then, there is a gap of size $\Delta\beta-2V_s$, between bands $\beta_2$ and $\beta_3$ for $\Delta\beta\geqslant2V_s$, and between bands $\beta_1$ and $\beta_2$ for $\Delta\beta<2V_s$, which does not depend on the interaction between $s$ and $p$ modes. On the contrary, the gap between bands $\beta_3$ and $\beta_4$ changes depending on the curvature of band $\beta_4$, what strongly depends on coupling $V_{sp}$ as shown in Fig.~\ref{fig2}. This change in the curvature necessarily implies that $\beta_4$ must be flat at some specific value of $V_{sp}$. By demanding that $\beta_4(0)=\beta_4(\pi)$, we obtain the following FB condition
\[
V_{sp}^{FB}\equiv V_s\sqrt{1+\frac{\Delta\beta}{2V_s}}\ .
\]
This mathematical relation is physical and experimentally possible due to the fact that $V_{sp}>V_s$, as expected considering the $s$ and $p$ mode profiles. If we fix coupling $V_s$, $V_{sp}^{FB}$ grows monotonically as a function of detuning $\Delta\beta$. Once this condition is fulfilled, the fourth band becomes completely flat with a value $$\beta_4=\Delta\beta +2V_s,$$ as shown in Figs.~\ref{fig3}(a) and (b) by a straight horizontal full orange line. We look for the eigenmode profile at this new FB condition. We assume a center site $n_0$ and an arbitrary amplitude $A$, obtaining that
\begin{eqnarray*}
u_{n} = A \delta_{n,n_0},\ v_{n} = \left(\frac{A}{2}\right)\left(\delta_{n,n_0}+\delta_{n,n_{0}-1}\right),\\ 
x_{n}=0,\ w_{n} = \left(\frac{V_s A}{2V_{sp}^{FB}}\right) \left(\delta_{n,n_0}-\delta_{n,n_{0}-1}\right).
\end{eqnarray*}
This profile is composed of both, $s$ and $p$, modes simultaneously and a sketch of it, on an effective ribbon lattice, is presented in Fig.~\ref{fig3}(c). We observe that the dipolar mode is smaller in amplitude with a factor $\sim 0.4$, for the parameters used in this figure. As coupling $V_{sp}^{FB}$ is larger than $V_s$, the mode amplitudes are compensated in order to satisfy a FB localization condition, which relies on destructive interference at specific connector sites~\cite{FBluis}. Superposed $s$ and $p$ mode amplitudes give the FB mode intensity profile sketched in Fig.~\ref{fig3}(d), for the 1D binary lattice system. The amplitudes beside the center show a shifted intensity with respect to the center of the waveguide, as expected from the superposition of fundamental and dipolar profiles at those sites. As a consequence, this localized state is very localized in space and perfectly compact.

A study of the transport in this lattice, performed by exciting a single vertical bulk site only (a delta-like input condition) would show a transition between dispersion (transport), localization (insulation), and transport again, while varying parameter $V_{sp}$. Localization would occur close to the FB condition $V_{sp}^{FB}$, while transport would manifest away this value. This behavior is quite similar to the one found for Sawtooth lattices~\cite{OLsaw}, where a FB is formed only for a very specific condition between coupling constants. Therefore, our simple 1D binary model could show an insulator transition when coupling interaction $V_{sp}/V_s$ is varied along the experiment. This could be demonstrated by fabricating several lattices having different refractive index profiles or directly shown by varying the temperature of a single binary lattice to achieve a tuning on propagation constants~\cite{bloch}.

\section{Edge states}

\begin{figure}[t!]\centering
\includegraphics[width=8.6cm]{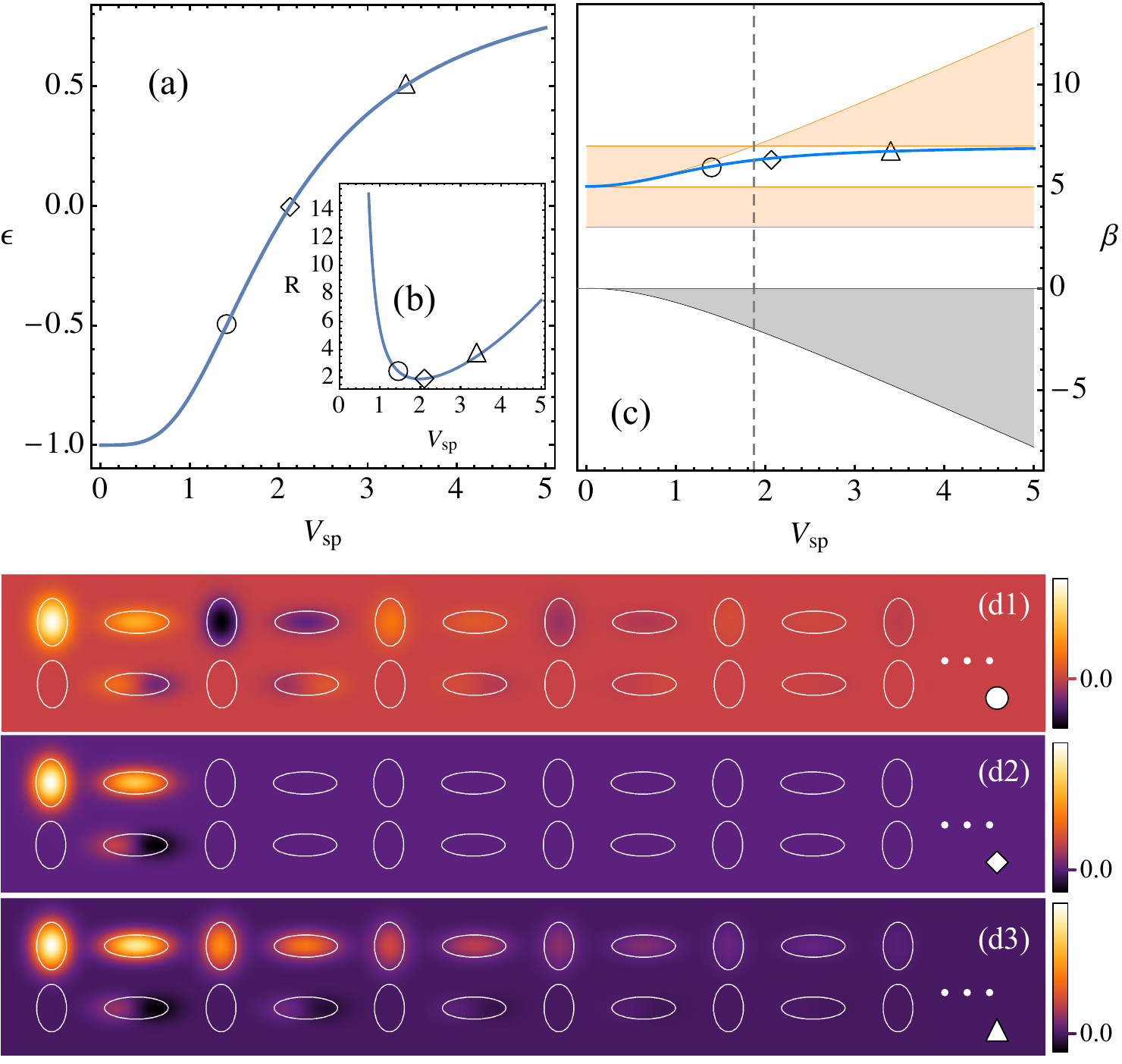}
\caption{(a) Decaying factor $\epsilon$, (b) participation ratio $R$, and (c) propagation constant $\beta_{e}$ versus coupling $V_{sp}$, for the edge mode (full blue line). Bands are plotted in (c) as shaded regions. (d1)--(d3) Effective edge mode amplitude profiles for $V_{sp}:\sqrt{2},\ V_{sp}^{ce}=2.13,\ 3.43$, respectively, labeled by a circle, a diamond and a triangle in (a)--(d). $\Delta\beta=5$ and $V_s=1$.}\label{fig4}
\end{figure}

When solving the eigenvalue problem (\ref{ep}), we look for solutions assuming an infinite lattice. Therefore, finite size effects, as for example linear edge modes, will not appear explicitly~\cite{OLsaw}. However, by numerically diagonalizing a finite lattice system, we find that an edge with a vertically oriented waveguide generates an exponentially decaying eigenmode, while a horizontal edge waveguide does not. In order to investigate this edge state, we consider a vertical waveguide at site $n=1$ and assume the following ansatz~\cite{rep1,rep2,OLsaw}
\begin{eqnarray*}
\{u_{n},v_{n},x_{n},w_{n}\}(z) = \{A,B,C,D\} \epsilon^{n-1}e^{i \beta_e z},
\end{eqnarray*}
for $n\geqslant1$, with $|\epsilon|<1$ (which implies an exponentially decaying state). $A,B$ and $D$ correspond to the amplitudes of this mode to be determined by solving a set of coupled equations. We assume a zero amplitude for mode $x_n$ ($C=0$), due to the no interaction of this mode with the rest of the system. (By taking $x_n\neq 0$ the frequency of this amplitude will be just zero, what not necessarily coincides with the frequency of the edge mode $\beta_e$. Additionally, there is always a perfectly localized edge state $x_n=C\delta_{n,1}$, as a trivial FB solution.) 
We insert this ansatz into model (\ref{eqs}) and write the equations for sites $n=1$ and $n=2$. We obtain two sets of three coupled equations, where the second set is recursively repeated for $n> 2$, what validates the proposed ansatz. These equations are the followings:
\begin{eqnarray}\label{eqeps1}
\beta_{e} A &=&\beta_s A+V_s B+V_{sp} D  ,\\
\beta_{e} B &=&\beta_s B+V_s A(1+\epsilon)  ,\nonumber\\ 
\beta_{e} D &=&\beta_p D+V_{sp} A(1-\epsilon), \nonumber
\end{eqnarray}
and
\begin{eqnarray}\label{eqeps2}
\beta_{e} A \epsilon &=&\beta_s A\epsilon+V_s B(1+\epsilon)+V_{sp} D(\epsilon-1)\  ,\\
\beta_{e} B \epsilon &=&\beta_s B\epsilon+V_s A\epsilon(1+\epsilon)\  ,\nonumber\\ 
\beta_{e} D \epsilon &=&\beta_p D\epsilon+V_{sp} A\epsilon(1-\epsilon)\ . \nonumber
\end{eqnarray}
By applying some algebra to equations (\ref{eqeps1}) and (\ref{eqeps2}), we obtain that
\[
\left(\frac{D}{A}\right)=\left(\frac{V_s}{V_{sp}}\right)\gamma,\ \ \epsilon=2\gamma^2-1,\ \ \beta_{e}=\beta_s+2V_s \gamma,
\]
with
\[
\gamma\equiv \left(\frac{B}{A}\right)=\frac{\sqrt{\Delta\beta^2 V_s^2+16 V_{sp}^2(V_s^2+V_{sp}^2)}-\Delta\beta V_s}{4(V_s^2+V_{sp}^2)}\ .
\]
This expression satisfies that $0<\gamma<1$, what implies that $-1<\epsilon<1$; i.e., this edge state is exponentially localized at the surface when this surface has a vertically oriented first waveguide. Additionally, as $V_{sp}>V_s$ then $(D/A)<(B/A)$; therefore, the edge localization is reinforced with a decreasing profile into the bulk of the system. In order to study the effective spatial size of these edge states, we compute an effective participation ratio, defined as $R\equiv [\sum_n(|u_n|^2+|v_n|^2+|x_n|^2+|w_n|^2)]^2/\sum_n(|u_n|^4+|v_n|^4+|x_n|^4+|w_n|^4)$, obtaining $$R=\frac{\left[1+\gamma^2+(V_s/V_{sp})^2\gamma^2\right]^2(1+\epsilon^2)}{\left[1+\gamma^4+(V_s/V_{sp})^4 \gamma^4 \right]\hspace{0.3cm} (1-\epsilon^2)}\ .$$ In order to characterize these states, we plot the decaying factor $\epsilon$, the participation ratio $R$ and the frequency $\beta_{e}$ versus the coupling $V_{sp}$ in Figs.~\ref{fig4}(a)--(c), respectively, for some specific values of $\Delta\beta$ and $V_s$ (the same phenomenology persists for different values).
%
%
First of all, we observe that when coupling $V_{sp}\rightarrow 0$, $B$ and $D$ goes to zero as well and the edge state bifurcates at the band center of a standard 1D lattice~\cite{rep1,rep2}, with $\epsilon\rightarrow -1$, $R\rightarrow\infty$, and $\beta_e\rightarrow\beta_s$. This state coincides with the $\pi/2$ linear mode of a standard 1D lattice and have an effective spatial profile of only $s$-mode amplitudes. Once we increase the coupling $V_{sp}$, we observe that the decaying factor $\epsilon$ decreases in magnitude, being for example $-0.5$ for $V_{sp}\approx \sqrt{2}$. In this case, the edge mode has a staggered profile every two sites, as shown in Fig.~\ref{fig4}(d1). After this, we obtain a perfectly localized edge state with an exactly zero tail ($\epsilon=0$), as Fig.~\ref{fig4}(d2) shows. This compact edge mode is obtained for the condition$$V_{sp}^{ce}\equiv V_s\sqrt{1+\frac{\Delta\beta}{\sqrt{2}V_s}}>V_{sp}^{FB}\ .$$ This state is similar to the edge mode found in Sawtooth lattices~\cite{OLsaw}, when different amplitudes destructively interfere at the connector sites of the lattice, in this case at the second vertically oriented waveguide. Although this profile corresponds to a perfectly compact edge state, which occupies only two sites of the lattice, with a mixed $s$-$p$ profile, it is not the most localized edge state on this 1D binary system. In fact, for the parameters considered in Fig.~\ref{fig4}, the perfectly compact edge state at $V_{sp}^{ce}=2.13$ has a participation ratio of $R=1.92$, while the minimum participation ratio $R=1.89$ occurs for $V_{sp}=1.98$.

After this regime, the decaying factor starts to grow slowly and profiles become completely unstaggered in their phase structure, as the example shown in Fig.~\ref{fig4}(d3) for $\epsilon\approx 0.5$. By a further increment of $V_{sp}$, $\epsilon$ slowly tends to its upper bound $1$, implying a smooth increment of $R$. The propagation constant $\beta_e$ slowly tends to $\Delta\beta+2V_s$, which coincides with the bottom of band $\beta_4$. It is important to notice that the FB condition at $V_{sp}^{FB}=1.87$ [see dashed vertical line in Fig.~\ref{fig4}(c)] produces an exchange on band $\beta_4$, in which the fundamental unstaggered mode passes from being at the top of the band for $V_{sp}=0$, to be at the band bottom for $V_{sp}>V_{sp}^{FB}$, as shown in Fig.~\ref{fig4}(c). Finally, a larger $V_{sp}$ coefficient implies that $\delta\rightarrow 1$, with $B/A\rightarrow 1$ and $D/A\rightarrow 0$. Therefore, the lattice effectively transforms into a standard 1D system, with a homogenous spatial profile of $s$-mode amplitudes only, which coincides with a standard unstaggered fundamental mode. Here, although $V_{sp}$ is large compared to $V_s$, there is a consecutive cancellation of dipolar amplitudes $D$, due to the alternated sign of this coupling interaction.

Both limits ($V_{sp}\rightarrow 0,\infty$) gives us an extended mode which coincides with the $\beta_4$-band modes of standard 1D lattices, where no surface states exist without distorting the lattice border~\cite{1DOL,1DPRL,1DOLKip}. As we observe in Fig.~\ref{fig4}(c), $\beta_{e}$ is only allowed to exist in the region $\{\Delta\beta,\Delta\beta+2V_s\}$, and edge modes behave quite similar to the one found in a Sawtooth lattice~\cite{OLsaw}, including the band twist at the FB critical parameter. As our edge modes do not appear due to any lattice perturbation at the border of the system~\cite{tamm,1DOL,1DPRL,1DOLKip}, but due to a crossover (twist) of the fourth linear band, the found edge modes are simply classified as Shockley-surface states~\cite{shock1,shock2,shock3}.

\section{$\Delta\beta=0$ limit}

\begin{figure}[t!]\centering
\includegraphics[width=8.6cm]{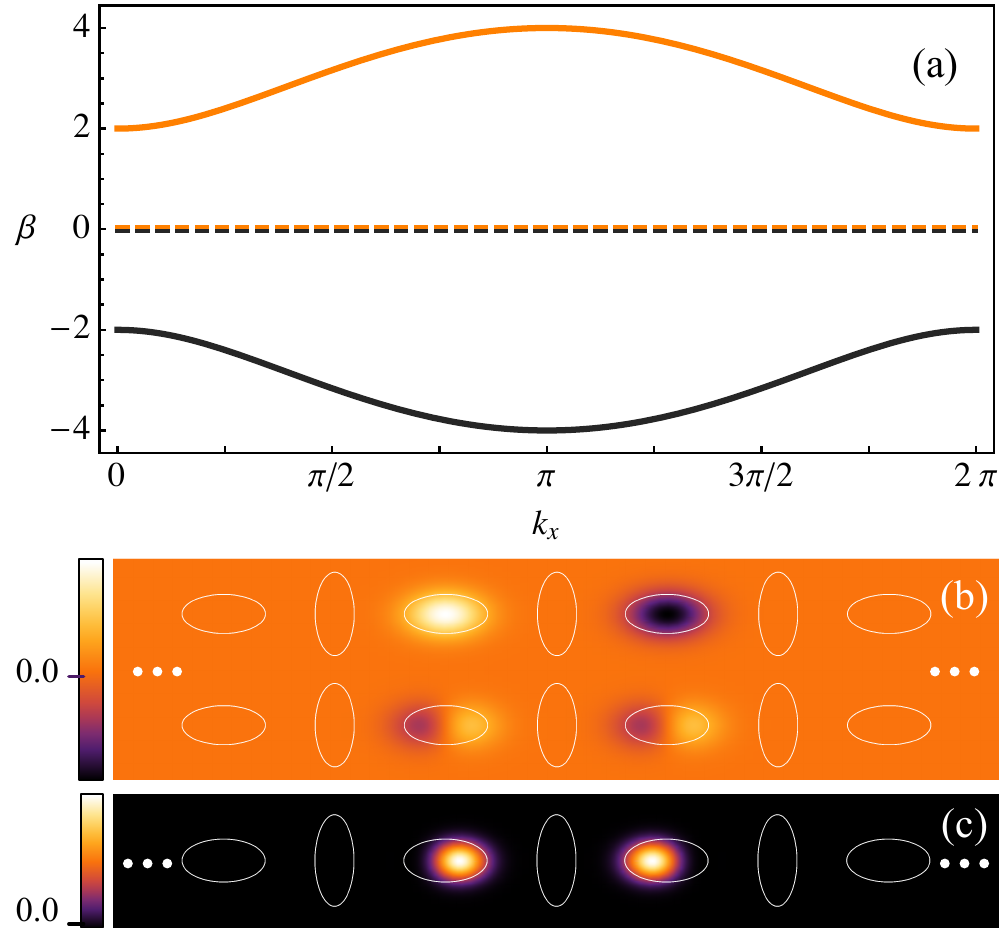}
\caption{(a) Linear spectrum for $\Delta\beta=0$, $V_s=1$, and $V_{sp}=2.0$. Full black, dashed black, dashed orange, and full orange correspond to $\beta_1$, $\beta_2$, $\beta_3$ and $\beta_4$ bands, respectively. (b) Effective amplitude and (c) intensity FB ($\beta_3$) mode profiles.}\label{fig5}
\end{figure}
%
Although the case $\Delta\beta=0$ corresponds to a non-physical solution in our photonic system~\cite{snyder,gloge,graphenNJP}, it becomes interesting to analyze it due to its phenomenology. By applying this condition, the eigenvalue problem (\ref{ep}) gives four simple solutions: $$\beta=0,0,\pm2\sqrt{V_s^2 \cos^2\bar{k}+V_{sp}^2 \sin^2\bar{k}}\ .$$ We plot the linear spectrum in Fig.~\ref{fig5}(a) and observe two opposite dispersive bands, showing a particle-hole symmetry~\cite{phs} in which for any value of $k_x$ there are two eigenfrequencies $\pm\beta(k)$. Additionally, we found two flat bands at exactly the same frequency $\beta_2=\beta_3=0$. The first one is the previously found trivial FB $\beta_2=0$, which consists on single-site vertical dipolar states. The second FB is generated by a combination of horizontal $s$ and $p$ modes, whose mode profiles consist on four amplitudes different to zero, having the following structure:
\begin{eqnarray*}
u_{n} = 0,\ v_{n} = A\left(\delta_{n,n_0}-\delta_{n,n_{0}+1}\right),\\ 
x_{n}=0,\ w_{n} = -\left(\frac{V_s}{V_{sp}}\right)A\left(\delta_{n,n_0}+\delta_{n,n_{0}+1}\right).
\end{eqnarray*}
This amplitude profile is sketched in Fig.~\ref{fig5}(b), with the corresponding intensity profile shown in Fig.~\ref{fig5}(c). We observe how a perfect cancellation of amplitudes, at connector sites, allows the formation of a highly localized pattern, which has only two sites different to zero. In terms of localization area, this state is comparable with the one found for diamond lattices~\cite{OLdiamond}, which is the most localized FB state observed experimentally up to now.

\section{Conclusions}

In conclusion, we have proposed a new model for the study of localization and transport of light in photonic lattices. In particular, our model consists on a rather simple 1D lattice having alternated orientation of elliptical waveguides. We found that, by assuming two -- $s$ and $p$ -- modes per site, a quasi-1D effective ribbon system emerges, which describes the light dynamics on this lattice. We found that there is always a FB on this system, which corresponds to vertically isolated dipolar states. These FB modes occupy a single site only being, therefore, the most localized FB states of any lattice configuration. By fulfilling a specific relation between lattice parameters, we found that a second non-trivial FB appears, which is composed of an hybridized state with $s$ and $p$ modes excited simultaneously. These FB states occupy only three lattice sites, with a rapidly decaying and perfectly compact profile. By investigating the edges of this lattice, we found that Shockley-like surface states exist on the system for edges having vertically oriented waveguides. We obtained an analytical expression for them and found that they could show different properties depending on the lattice parameters. At the end, we explored the case $\Delta\beta=0$ and found two dispersive and two flat bands for this binary 1D system. 

We believe that our simple model could show interesting features for non-diffractive image transmission applications as well as for presenting different transport properties depending on the input condition. For example, by exciting a vertically oriented waveguide with a fundamental state we would simply observe transport, while using a dipolar excitation would produce perfect localization, without the need of applying any extra interaction. This could be useful to excite two completely different states on the system, which could be of interest on optical signal processing.

\section*{Acknowledgements}

Authors acknowledge B. Real for useful comments and discussions in the initial part of this work. This work was supported in part by Programa ICM Millennium Institute for Research in Optics (MIRO) and FONDECYT Grant No.1151444.

\end{document}